\documentclass[a4paper,10pt]{article}
\pdfoutput=1
\usepackage{graphicx}
\usepackage{fancyheadings}
\begin{document}
\noindent
{\it 12th Hel.A.S Conference}\\
\noindent
{\it Thessaloniki, 28 June - 2 July, 2015}\\
\noindent
%
%
CONTRIBUTED LECTURE\\
\noindent
\underline{~~~~~~~~~~~~~~~~~~~~~~~~~~~~~~~~
~~~~~~~~}
\vskip 1cm
%
%
\begin{center}
{\Large\bf
Highly eccentric exoplanets trapped in mean-motion resonances
}
\vskip 0.5cm
%
%
{\it
K. I. Antoniadou and G. Voyatzis
}\\
%
%
Section of Astrophysics, Astronomy and Mechanics, Department of Physics, \\Aristotle University of Thessaloniki, 54124, Greece
\end{center}
\vskip 0.7cm
%
%
\noindent
{\bf Abstract: }
We herein utilize the general three-body problem (GTBP) as a model, in order to simulate resonant systems consisting of a star and two planets, where at least one of them is highly eccentric. We study them in terms of their long-term stability, via the construction of maps of dynamical stability and the computation of the corresponding families of periodic orbits. We identify the way their survival is connected with the regions of regular motion in phase space, which, in turn, were created by stable resonant periodic orbits in their vicinity. Consequently, a phase protection mechanism is provided and the planets avoid close encounters and collisions even on long timescales. We apply our methodology to the extrasolar system HD 82943.


\section{Introduction}
The number of known extrasolar planets has increased considerably over the last years and the discovery of these systems has raised many interesting questions e.g. about their formation, composition, habitability etc. Today, about 500 systems seem to consist of more than two planets and in many of them, the gravitational interaction between the planets is not negligible. Therefore, interesting questions about their dynamical evolution and stability are raised, too. 

\begin{figure}[bh]
\centering
\includegraphics[width=14cm]{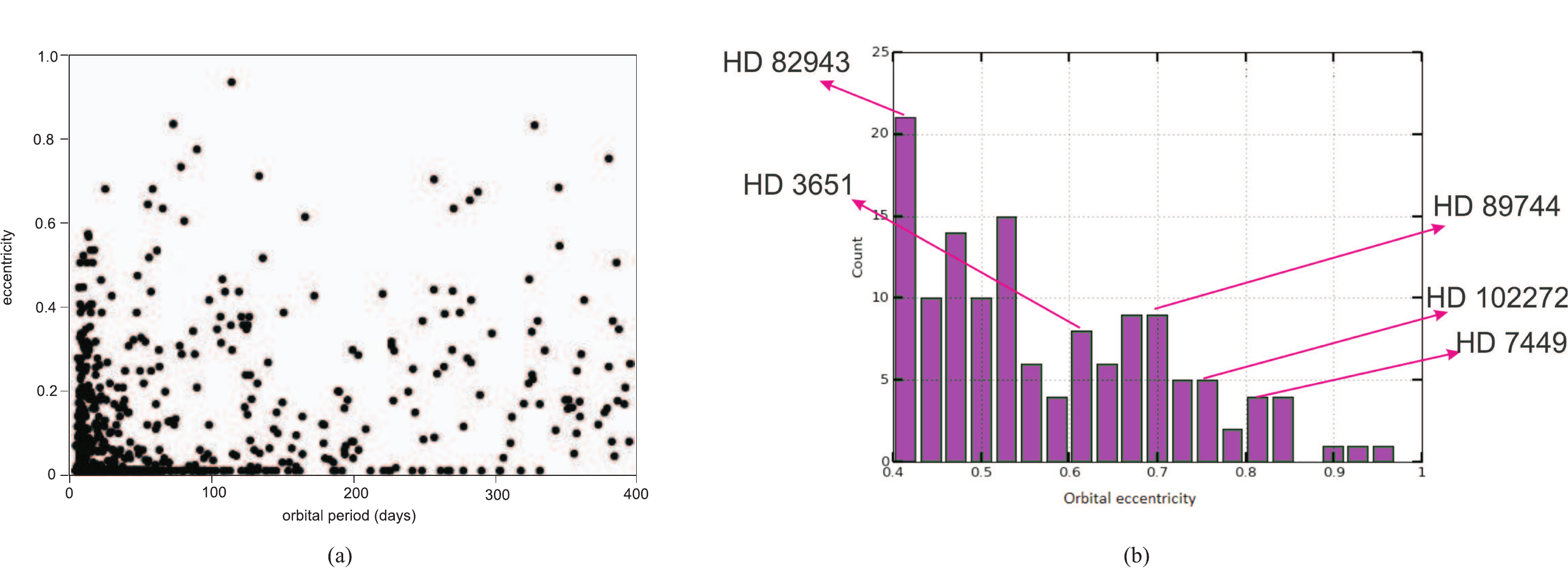}
\caption{Distribution of eccentricities of exoplanets (data from {\em exoplanet.eu})}
\label{Fg1}
\end{figure}

The  long-term stability of multi-planet systems becomes very interesting when eccentricities are quite high. In this case, the almost Keplerian elliptical orbits can be close to each other or intersect and, therefore, mutual gravitational interactions may be very strong. A mechanism that prevents the close encounter of planets is the resonant phase protection. When planets revolve so that their mean motion ratio is almost rational $n_1/n_2\approx p/q$, $p,q\in N$, i.e. we have a mean motion resonance, we can determine initial conditions of the planets, for which, although the orbits are very close to each other, the planets avoid coming close to one another and, therefore, their mutual interactions are weak. Subsequently, planetary systems of high eccentric orbits can only survive if they are trapped in a resonance and have the appropriate phases. 

According to the data provided by the extrasolar planets encyclopedia, many exoplanets have large eccentricities. The distribution of the eccentricities with respect to the orbital period is given in Fig. \ref{Fg1}(a). An important question is whether such eccentric planets can have companions \cite{Wittenmyer}. Nevertheless, according to the analysis of radial velocity data, multiple systems with eccentric planets seem to exist e.g. the systems HD 82943, HD 3651, HD 7449, HD 107222 and HD 89744 (see Fig. \ref{Fg1}(b)). Data from {\em Kepler} also indicate many planets with high eccentric orbits \cite{Kane}. In this paper, we present a methodology for obtaining stable orbits in two-planet systems  where, at least, one of the planets revolves in a highly eccentric orbit. The main aspects of two-planet systems dynamics are presented in \cite{Mitch08}.
           
\section{Methodology}
We model a two-planet system with the general planar three body problem consisting of a very massive star with mass $m_0$ and two planets with masses $m_1$ and $m_2$ (the inner and the outer planet, respectively). By using the osculating orbital elements, the parameters that define explicitly the dynamical position of the system are the ratio of semimajor axes $\alpha=a_2/a_1$, the eccentricities $e_1$ and $e_2$, the angle of pericenters $\Delta\varpi=\varpi_2-\varpi_1$ and the initial relative position on their osculating ellipses, which can be given by the resonant angle $\theta=p\lambda_2-q \lambda1-(p-q)\varpi_1$ \cite{AV2013}.  

As we mentioned above, highly eccentric orbits of long-term stability can exist in resonant regions. The centers of such regions (exact resonances) are periodic orbits of the three body problem in a rotating frame of reference \cite{Hadjidem06}. Resonant periodic orbits are not isolated, but they form monoparametric families in phase space. For the case of symmetric periodic orbits $\Delta\varpi$ and $\theta$ are constant and take the values 0 or $\pi$. This means that the planetary orbits are aligned or anti-aligned and the planets are in conjuction. These families can be represented by characteristic curves in the plane $e_1-e_2$. However, along such curves $\alpha$ varies slightly. 

\begin{figure}[bt]
\centering
\includegraphics[width=13cm]{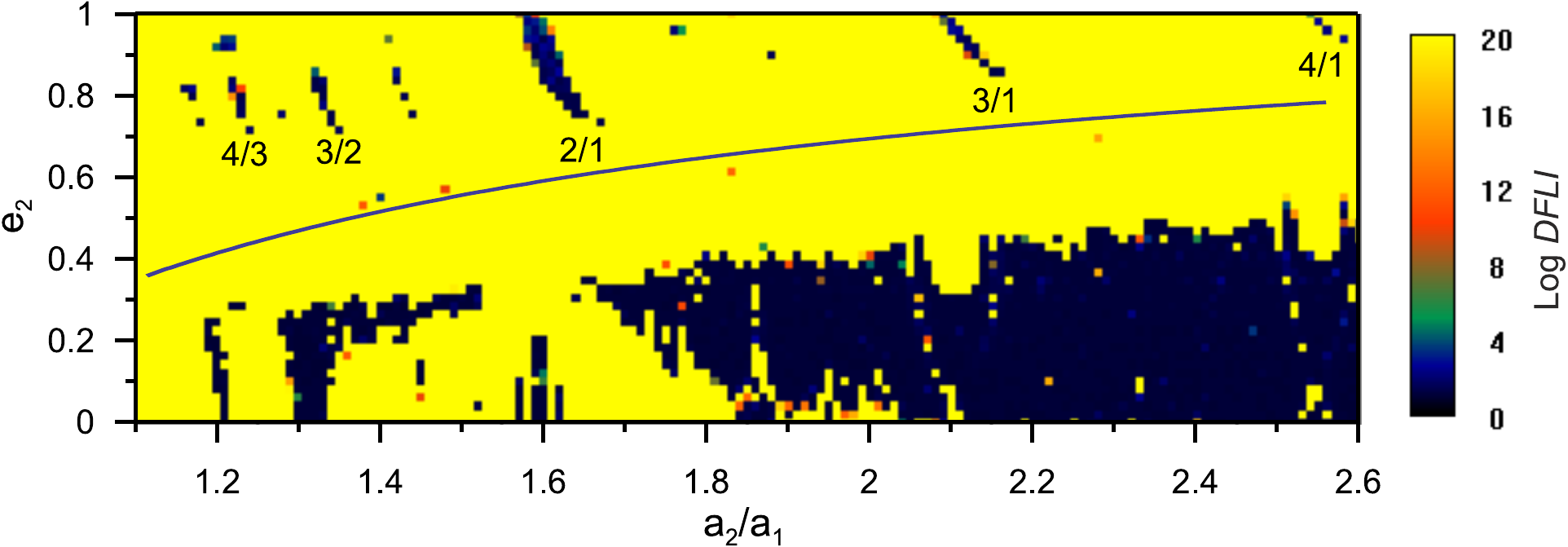}
\caption{A dynamical map in the plane $\alpha - e_2$ for Jupiter-Saturn masses ($e_1=0.3, \varpi_1=\varpi_2=0$, $M_1=M_2=0$, where $M_1$, $M_2$ are the mean anomalies). The corresponding resonances for the highly eccentric islands are indicated. Gray curve shows the collision line - above which the planetary orbits intersect.}
\label{Fg2}
\end{figure}

With the term ``stability'' we refer to orbits that evolve regularly, namely they are quasiperiodic orbits winding invariant tori in phase space. Chaotic orbits may also survive for long time integrations, however, the majority of them terminates at a collision or escape of one of the planets. Chaos indicators can be used for detecting the chaotic nature of orbits in relatively short time intervals (e.g. see \cite{Sandor}).  In our study, we use the {\em de-trented FLI} defined as the function
$$
DFLI(t)=sup_t (|\mathbf{\xi}(t)|/t),
$$
where $\mathbf{\xi}$ is the deviation vector of the orbit, which is evaluated along the numerical integration of the orbit simultaneously with the integration of the linearized system. Dynamical maps of stability can be computed by considering planar grids of initial conditions, numerical integration of orbits and computation of  the value $DFLI=DFLI(t_{max})$ \cite{Voyatzis2008}. An example is shown in Fig. \ref{Fg2}, where a map on the plane $\alpha - e_2$ is given by fixing $e_1=0.3$, $\Delta\varpi=\theta=0^\circ$ and setting the masses $m_1=0.001$, $m_2=0.0003$ (Jupiter - Saturn planets). Integrations performed for 200Ky (with $a_1=1$AU). Dark colors ($DFLI<10^4$) correspond to long-term stable orbits, while yellow color indicates strongly chaotic orbits with close encounters in relatively short time intervals. A notable main property of the DFLI is the sharp distinction between chaos and order. By obtaining a large value e.g. $DFLI>10^{20}$ we can stop the integration and classify the orbit as chaotic.  

The dynamical map shows that for $e_2<0.4$ (bellow the collision line) there exist many regions of stable motion. However, above the collision line, only distinct and small (but not negligible) stable regions are detected. These regions of initial conditions provide resonant motion and each one is built around a particular family of periodic orbits.

\begin{figure}[tb]
\centering
\includegraphics[width=10cm]{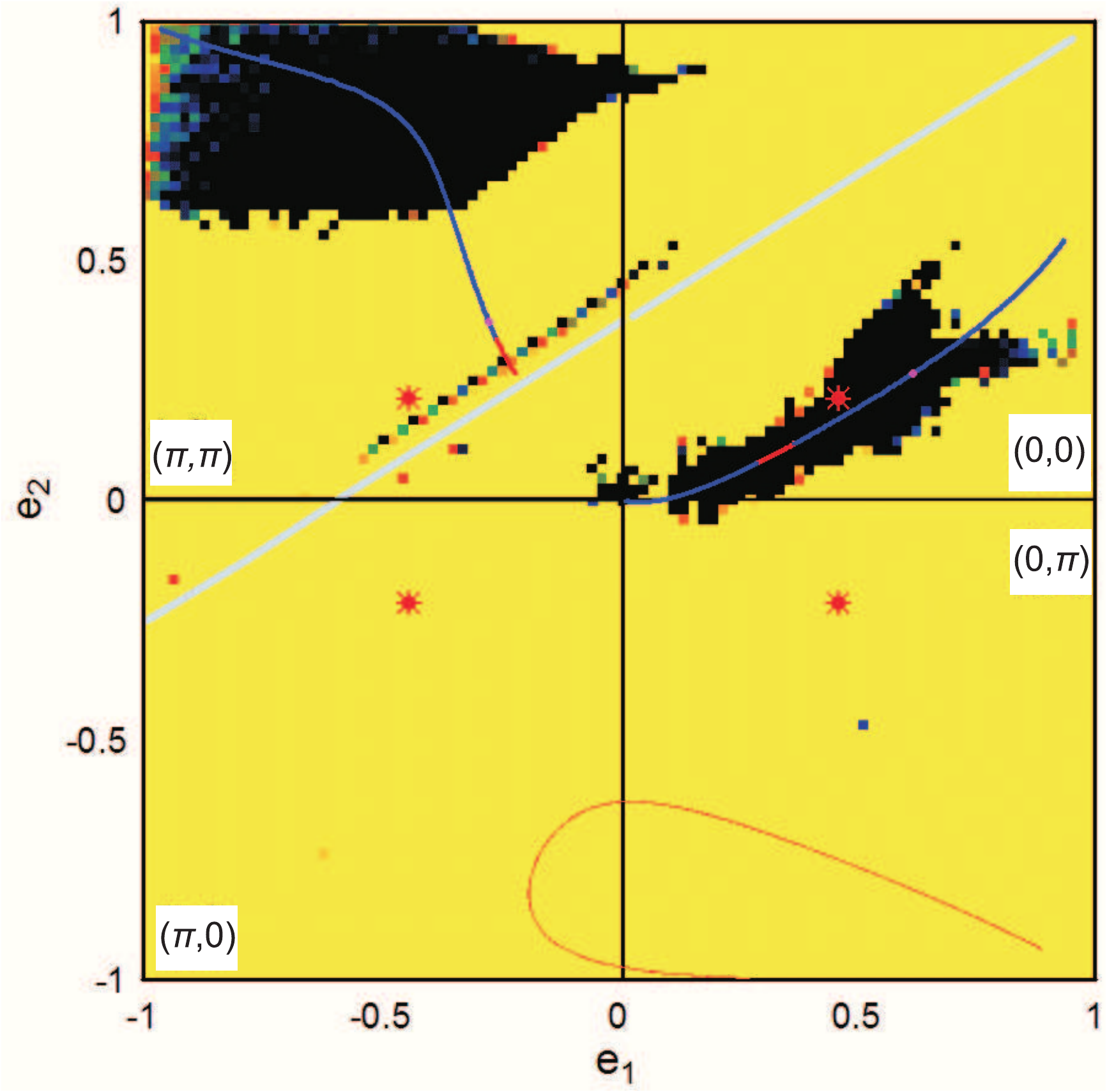}
\caption{Dynamical map for the 2/1 resonant motion and for the masses of the planets HD 82943b,c. Each quadrant corresponds to the indicated configuration $(\theta, \Delta\varpi)$. Bold blue and red curves are the the stable and unstable families of periodic orbits, respectively. The gray bold line is the collision line.}
\label{Fg3}
\end{figure}

\section{An application to HD 82943 extrasolar system.} 
The suggested parameters for the position of the system HD 82943 are \cite{Xianyu}
$$
m_1/m_0=0.0041, m_2/m_0=0.0042, a_1=0.74, a_2=1.18, e_1=0.43,  e_2=0.2, \Delta\varpi=25^\circ, \theta=-2^\circ.
$$
The above values indicate that the system should be trapped in 2/1 resonance and close to the symmetric configuration $\Delta\varpi=\theta=0^\circ$. For the particular masses we computed the main families of symmetric periodic orbits at the initial configurations $(\theta,\Delta\varpi)=(0,0)$, $(0,\pi)$, $(\pi,0)$, $(\pi,\pi)$. For the same configurations we computed the dynamical maps at the plane $e_1-e_2$ with $\alpha\approx 1.59$. Each map provides a quadrant of the global map presented in Fig. \ref{Fg3}. The $e_1-e_2$ characteristic curves of the families of periodic orbits are also presented. Bold blue segments indicate stable periodic orbits, while red ones consist of unstable periodic orbits. It is clearly shown that the configurations $(0,\pi)$ and $(\pi,0)$ do not support any stable orbits and families of stable periodic orbits do not exist. Instead, the presence of the stable family of periodic orbits for $(\theta,\Delta\varpi)=(\pi,\pi)$ is accompanied by a broad region of initial conditions of regular orbits. Wholly this region is located above the collision line and stable orbits can be found even for $e_i\approx 1$.   

\begin{figure}[htb]
\centering
\includegraphics[width=15cm]{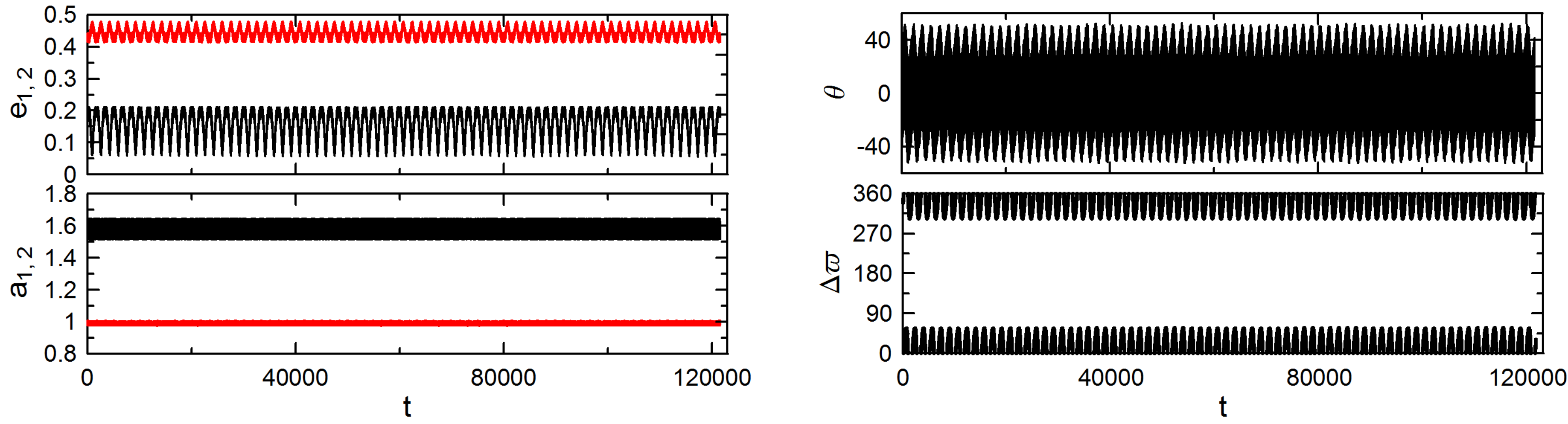}
\caption{The evolution of orbital elements of the extrasolar system HD 82943.}
\label{Fg4}
\end{figure}

The position of HD 82943 in the map is indicated by the red-star like symbol. We obtain that the family of periodic orbits in the $(0,0)$ configuration is surrounded by a region of stability. Starting with initial conditions in this region, the evolution of orbital elements is regular, and, particularly, the angles $\theta$ and $\Delta\varpi$ librate around $0^\circ$. The system HD 82943 is located in this region of stability. Starting with the initial conditions, given above, the orbital elements evolve as it is shown in Fig. \ref{Fg4}, and this picture is expected to remain unaltered for many Gyears.  

\vskip 0.5cm
\noindent
{\bf Acknowledgements:} This research has been co-financed by the European Union (European Social Fund - ESF) and Greek national funds through the Operational Program ``Education and Lifelong Learning'' of the National Strategic Reference Framework (NSRF) - Research Funding Program: Thales. Investing in knowledge society through the European Social Fund.

\end{document}